\documentclass[english]{revtex4-1}
\usepackage[T1]{fontenc}
\usepackage[latin9]{inputenc}
\usepackage{xcolor}
\usepackage{babel}
\usepackage{bm}
\usepackage{amsbsy}
\usepackage{amstext}
\usepackage{amssymb}
\usepackage{graphicx}
\usepackage{esint}
\usepackage[pdfusetitle,bookmarks=true,bookmarksnumbered=false, bookmarksopen=false,breaklinks=false,pdfborder={0 0 1},backref=false,colorlinks=true]{hyperref}
\hypersetup{pdfborderstyle=,colorlinks,citecolor=blue,linkcolor=blue}


\begin{document}
\title{Proof of bulk-edge correspondence for band topology by Toeplitz algebra}
\author{Zixian Zhou}
\author{Liang-Liang Wan}
\email{wanliangliang@sztu.edu.cn}
\affiliation{Shenzhen Key Laboratory of Ultraintense Laser and Advanced Material Technology, Center for Intense Laser Application Technology, and College of Engineering Physics, Shenzhen Technology University, Shenzhen 518118, China}

\begin{abstract}
We rigorously yet concisely prove the bulk-edge correspondence for general $d$-dimensional ($d$D) topological insulators in complex Altland-Zirnbauer classes, which states that the bulk topological number equals to the edge-mode index. Specifically, an essential formula is discovered that links the quantity expressed by Toeplitz algebra, i.e., hopping terms on the lattice with an edge, to the Fourier series on the bulk Brillouin zone. We then apply it to chiral models and utilize exterior differential calculations, instead of the sophisticated \emph{K}-theory, to show that the winding number of bulk system equals to the Fredholm index of 1D edge Hamiltonian, or to the sum of edge winding numbers for higher odd dimensions. Moreover, this result is inherited to the even-dimensional Chern insulators as each of them can be mapped to an odd-dimensional chiral model. It is revealed that the Chern number of bulk system is identical to the spectral flow of 2D edge Hamiltonian, or to the negative sum of edge Chern numbers for higher even dimensions. Our methods and conclusions are friendly to physicists and could be easily extended to other physical scenarios.
\end{abstract}
\maketitle

\section{Introduction}

Topology plays an increasingly important role in characterizing states of matter, especially the lattice band systems. Apart from the well known fermionic topological insulators and superconductors in Altland-Zirnbauer (AZ) classification\,\citep{Hasan2010RevModPhys,Qi2011RevModPhys,Chiu2016RevModPhys}, topological bands are also found and classified in quadratic-bosonic\,\citep{Lu2018arXiv,Xu2020PRB,Zhou2020JPA} and even non-Hermitian systems\,\citep{Gong2018PhysRevX,Kawabata2019PRX}. In all these systems, there is an iconic phenomenon called bulk-edge
correspondence (BEC)\,\citep{Laughlin1981PRB,Hatsugai1993PRL}, that is, when the lattice has an edge, edge states would occur within the band gap of the bulk system and exhibit a topological behavior related to the bulk topology but irrespective to other physical details. This striking phenomenon not only implicates a profound index theorem in mathematical physics, but also associates with many practical effects like quantized Hall conductivity\,\citep{Thouless1982PRL} and fractional statistics of Majorana edge modes \citep{Nayak2008RMP}. Therefore, a clear understanding of BEC becomes a fundamental problem.

Although the BEC in free-fermion systems has been studied by twisted equivariant \emph{K}-theory\,\citep{Kellendonk2002RIMP,Prodan2016}, it is too sophisticated for physicists and hard to extend to other physical systems. Besides, physical community has also paid many efforts to achieve concise proofs to various degrees of rigor and generality. The physical approaches include Green's functions\,\citep{Essin2011PRB}, Levinson's theorem in scattering theory\,\citep{Graf2013CMP} and Atiyah-Singer index theorem\,\citep{Fukui2012JPSJ}. Nevertheless, all these prior works have made approximations and have somewhat limitations. For instance, the approach of Green's function discards higher order expansion terms which are not infinitesimal; the approach of long-wave limit reduces the lattice Hamiltonian to a differential operator on a continuum system, which alters the essence of lattice. Thus, the key problem still exists: Is there a rigorous yet concise approach to prove the BEC for fermionic band systems, which can be easily transferred to other physical scenarios like bosonic and non-Hermitian systems?

This work answers yes to the problem. We use Toeplitz algebra and exterior differentiation instead of the sophisticated \emph{K}-theory to rigorously yet concisely prove the BEC for general dimensional topological insulators in classes A and AIII according to the AZ classification\,\citep{Chiu2016RevModPhys}. Specifically, the tight-binding models on the lattice with an edge are described by a polynomial of Toeplitz matrix that represents the hopping term of particles on the half infinite lattice. Via replacing the Toeplitz matrices by plane wave functions, the edge Hamiltonian (i.e., the Hamiltonian on the lattice with an edge) simply reduces to the Bloch Hamiltonian of the bulk. This is known as a homomorphism between the Toeplitz algebra generated by Toeplitz matrices and Fourier series. Neither the \emph{K}-theory approach utilizing exact sequences nor the former physical approaches avoiding Toeplitz algebra, here we discover a simple and powerful formula telling that the trace of commutator of Toeplitz algebra equals to an integral of Fourier series. This implies that a proper quantity formulated by edge Hamiltonian can be converted to that by bulk one. Therefore, our strategy seeks to define edge-mode indices explicitly by edge Hamiltonians and also write the bulk topological numbers by bulk Hamiltonians. With this task done, the equivalence of bulk and edge topological quantities can be easily proved by algebraic and calculus computations. In this way, a clear understanding of the BEC is provided for fermionic insulators, which is transferable to other physical systems.

Based on the above strategy, we firstly investigate the odd dimensional chiral models in class AIII, for their bulk topology is characterized by a winding number which already has an explicit expression by bulk Hamiltonian. Besides, the topology of edge modes is characterized by a Fredholm index in one dimension which is equivalent to the partition function of (square) edge Hamiltonian in zero temperature. In higher odd dimensions, we successfully define an edge winding number which can still be formulated by the edge Hamiltonian. Thus, our strategy is implemented. By applying our formula, we are able to prove that the bulk winding number equals to the edge-mode index, that is, to prove the BEC.

Next, we turn to prove the BEC for the even dimensional Chern insulators in class A whose bulk topology is characterized by a Chern number formulated by Berry curvature rather than Hamiltonian itself. Fortunately, each $d$-dimensional ($d$D) Chern insulator can be mapped to a $\left(d+1\right)$D chiral model by a simple construction. It turns out that the Chern number of the Chern insulator is identical to the winding number of the constructed chiral model up to a sign. The edge-mode index of the Chern insulator is also induced by that of the chiral model. Therefore, the BEC of Chern insulators is inherited from the former result of chiral models. We thus concisely prove the BEC for all the complex AZ classes in any dimension.

Our article is organized as follows. In Sec. \ref{sec:Insulators}, the edge and bulk Hamiltonians are given. In Sec. \ref{sec:Toeplitz}, the properties of Toeplitz algebra and the formula are revealed. In Sec. \ref{sec:chiral}, the BEC for odd dimensional chiral models is proved. Subsequently, in Sec. \ref{sec:Chern insulator}, the BEC for even dimensional Chern insulators is proved. In Sec. \ref{sec:conclusions}, conclusions are made.

\section{Insulators with and without boundary\label{sec:Insulators}}

We study the $d$D fermionic insulator described by a tight-binding model on an infinite crystal lattice with $N$ internal degrees of freedom (e.g., spins) in each unit cell. By denoting $\hat{c}_{\bm{r}}=\left(\hat{c}_{\bm{r}1},\cdots,\hat{c}_{\bm{r}N}\right)^{T}$ the column vector of fermionic annihilation operators at lattice site $\bm{r}=\left(r_{1},\ldots,r_{d}\right)$, we write down the many-body Hamiltonian 
\begin{equation}
\hat{H}=\sum_{\bm{r},\bm{r}^{\prime}}\hat{\bm{c}}_{\bm{r}}^{\dagger}h_{\bm{r-r^{\prime}}}\hat{\bm{c}}_{\bm{r}^{\prime}},
\end{equation}
where $h_{\bm{r-r^{\prime}}}\in M_{N}\left(\mathbb{C}\right)$ is an $N\times N$ matrix and describes the hopping of particle from site $\bm{r}^{\prime}$ to $\bm{r}$. These hopping terms are finite in physical models.

We firstly consider the half infinite lattice structured by $\mathbb{N}\times\mathbb{Z}^{d-1}$ which has an edge ($r_{1}\in\mathbb{N}$) along the $r_{1}$-direction and has no boundaries along the remaining dimensions ($\tilde{\bm{r}}=\left(r_{2},\cdots,r_{d}\right)\in\mathbb{Z}^{d-1}$). After the Fourier transformations $\hat{c}_{r_{1}}\left(\tilde{\bm{k}}\right)=\sum_{\tilde{\bm{r}}}\hat{c}_{\bm{r}}e^{-{\rm i}\tilde{\bm{k}}\cdot\tilde{\bm{r}}}$ with crystal momenta $\tilde{\bm{k}}=\left(k_{2},\ldots,k_{d}\right)\in T^{d-1}$ ($k_{i}\in\left[0,2\pi\right)$), the degrees of freedom $\tilde{\bm{k}}$ are decoupled in the Hamiltonian 
\begin{equation}
\hat{H}=\int\frac{\text{d}k_{2}\cdots\text{d}k_{d}}{\left(2\pi\right)^{d-1}}\sum_{r,r^{\prime}\in\mathbb{N}}\hat{c}_{r}^{\dagger}\left(\tilde{\bm{k}}\right)h_{r-r^{\prime}}\left(\tilde{\bm{k}}\right)\hat{c}_{r^{\prime}}\left(\tilde{\bm{k}}\right)
\end{equation}
with $h_{r_{1}}\left(\tilde{\bm{k}}\right)=\sum_{\tilde{\bm{r}}}h_{\bm{r}}e^{-{\rm i}\tilde{\bm{k}}\cdot\tilde{\bm{r}}}$. Here the associated Brillouin zone (BZ) is a $\left(d-1\right)$D torus $T^{d-1}$. Now the transformed Hamiltonian matrix $H_{rr^{\prime}}=h_{r-r^{\prime}}\left(\tilde{\bm{k}}\right)$ describes an effective 1D tight-binding Hamiltonian model on lattice $\mathbb{N}$, parameterized by $\tilde{\bm{k}}$. With the help of Toeplitz matrix 
\begin{equation}
T=\left(\begin{array}{cccc}
0\\
1 & 0\\
 & 1 & 0\\
 &  & \ddots & \ddots
\end{array}\right),\,T_{i,j}=\delta_{i-1,j},
\end{equation}
this Hamiltonian $H=\left(H_{rr^{\prime}}\right)$ is expressed as 
\begin{equation}
H\left(\tilde{\bm{k}}\right)=\sum_{r\geq0}h_{-r}\left(\tilde{\bm{k}}\right)\otimes\left(T^{\dagger}\right)^{r}+\sum_{r>0}h_{r}\left(\tilde{\bm{k}}\right)\otimes T^{r}.\label{eq:H_edge}
\end{equation}
Here $T$ and $T^{\dagger}$ describe the nearest-neighbor hopping of particles towards the infinity and the edge side, respectively. They satisfy identity 
\begin{equation}
T^{\dagger}T=I\label{eq:ident}
\end{equation}
with identity operator $I$ but $TT^{\dagger}\neq I$. In the follows, we are interested in the bound eigenstates of $H\left(\tilde{\boldsymbol{k}}\right)$ which are known as the edge modes. Hence we restrict $H\left(\tilde{\bm{k}}\right)$ to the Hilbert space $\mathbb{C}^{N}\otimes\ell^{2}\left(\mathbb{N}\right)$ with the scattering states excluded. Then $H\left(\tilde{\boldsymbol{k}}\right)$ becomes an effective Hamiltonian for the edge degrees of freedom, dubbed as edge Hamiltonian. Since the hopping terms are finite, $H\left(\tilde{\bm{k}}\right)$ is a bounded self-adjoint operator and is smooth regarding with $\tilde{\bm{k}}$.

We next consider the lattice structured by $\mathbb{Z}^{d}$ without boundaries, which is equivalent to the periodic boundary condition with infinite crystal volume. In this case, full-dimension Fourier transformations $\hat{c}\left(\bm{k}\right)=\sum_{\bm{r}}\hat{c}_{\bm{r}}e^{-{\rm i}\bm{k}\cdot\bm{r}}$ can be made with $\bm{k}=\left(k_{1},\ldots,k_{d}\right)\in T^{d}$ ($k_{i}\in\left[0,2\pi\right)$), and all the degrees of freedom are decoupled in the Hamiltonian 
\begin{equation}
\hat{H}=\int\frac{\text{d}k_{1}\cdots\text{d}k_{d}}{\left(2\pi\right)^{d}}\hat{c}^{\dagger}\left(\bm{k}\right)h\left(\bm{k}\right)\hat{c}\left(\bm{k}\right),
\end{equation}
in which the bulk Hamiltonian (i.e., Bloch Hamiltonian) is given by
\begin{equation}
h\left(\bm{k}\right)=\sum_{\bm{r}}h_{\bm{r}}e^{-{\rm i}\bm{k}\cdot\bm{r}}.\label{eq:h_bulk}
\end{equation}
The Hermiticity of $h\left(\bm{k}\right)$ stems from the many-body Hamiltonian $\hat{H}$. Here we assume that the bands are gapped at zero energy, i.e., $\det h\left(\bm{k}\right)\neq0$ for all $\bm{k}$.
This restriction is natural for the fermionic insulators and paves the road to define topological invariants for bulk Hamiltonians.

\section{Toeplitz algebra\label{sec:Toeplitz}}

The tight-binding model on the half infinite lattice, represented by Toeplitz matrices, does not have the complete crystal translation symmetry. As a result, these degrees of freedom cannot be decoupled by Fourier transformation, and many quantities like eigen-energies are difficult to compute. Nevertheless, there is still a homomorphism relation between Toeplitz matrices and Fourier series. Through this relation, some special quantities like topological invariant can be cleverly computed.

To be specific, $T$ generates a $C^{\ast}$-algebra (by addition, multiplication, $\mathbb{C}$-number multiplication, and Hermitian conjugation) called Toeplitz algebra $\mathbb{T}$ \citep{Arveson2001}. Each element $X\in\mathbb{T}$, as an operator on the Hilbert space $\ell^{2}\left(\mathbb{N}\right)$, is required to be bounded. Due to the identity Eq. (\ref{eq:ident}), each $X\in\mathbb{T}$ can be uniquely expanded as 
\begin{equation}
X=\sum_{m,n\in\mathbb{N}}x_{mn}T^{m}\left(T^{\dagger}\right)^{n},\,x_{mn}\in\mathbb{C}.
\end{equation}
On the other hand, the bounded functions on the 1D BZ $S^{1}$ also form a $C^{\ast}$-algebra $L^{\infty}\left(S^{1}\right)=\left\{ \sum_{n\in\mathbb{Z}}f_{n}e^{-{\rm i}nk}\right\}$ in which $\left\{ f_{n}\right\} \in\ell^{1}\left(\mathbb{Z}\right)$ are Fourier coefficients. There is a $C^{\ast}$-algebra \emph{homomorphism} $\varphi:\mathbb{T}\rightarrow L^{\infty}\left(S^{1}\right)$ given by the replacement $T\mapsto e^{-{\rm i}k}$, i.e., 
\begin{equation}
\varphi\left(\sum_{m,n\in\mathbb{N}}x_{mn}T^{m}\left(T^{\dagger}\right)^{n}\right)=\sum_{m,n\in\mathbb{N}}x_{mn}e^{-{\rm i}mk}e^{{\rm i}nk}.
\end{equation}
One can readily check the following homomorphism axioms (for $X,Y\in\mathbb{T}$ and $\mu,\nu\in\mathbb{C}$) 
\begin{eqnarray}
\varphi\left(\mu X+\nu Y\right) & = & \mu\varphi\left(X\right)+\nu\varphi\left(Y\right),\label{eq:linear}\\
\varphi\left(XY\right) & = & \varphi\left(X\right)\varphi\left(Y\right),\label{eq:mult}\\
\varphi\left(X^{\dagger}\right) & = & \left[\varphi\left(X\right)\right]^{\dagger},\label{eq:conju}
\end{eqnarray}
and see $\left[\varphi\left(T\right)\right]^{\dagger}\varphi\left(T\right)=1$ consistent with the identity Eq. (\ref{eq:ident}). After extending the algebraic homomorphism to $\varphi:M_{N}\left(\mathbb{C}\right)\otimes\mathbb{T}\rightarrow M_{N}\left(\mathbb{C}\right)\otimes L^{\infty}\left(S^{1}\right)$ and let $k=k_{1}$, one immediately finds that $\varphi$ maps the edge Hamiltonian Eq. (\ref{eq:H_edge}) to the bulk one Eq. (\ref{eq:h_bulk}), i.e., 
\begin{equation}
\varphi\left(H\left(\tilde{\bm{k}}\right)\right)=h\left(\bm{k}\right).\label{eq:homo}
\end{equation}

Based on this homomorphism, we put forward an important \emph{formula} that links the quantity expressed by Toeplitz matrices to that by Fourier series: For $X,Y\in M_{N}\left(\mathbb{C}\right)\otimes\mathbb{T}$, there is 
\begin{equation}
\text{Tr}\left[X,Y\right]=\frac{{\rm i}}{2\pi}\text{tr}\int_{S^{1}}\varphi\left(X\right)\text{d}\varphi\left(Y\right),\label{eq:formula}
\end{equation}
in which $\text{Tr}$ traces over the Hilbert space and $\text{tr}$ traces over finite dimensional space. This implies that $\text{Tr}\left[X,Y\right]$ is finite if $\varphi\left(X\right),\varphi\left(Y\right)\in C^{1}\left(S^{1}\right)$.

The proof is quite simple. Due to the bi-linearity, we just need to verify the basis of $\mathbb{T}$ satisfying Eq. (\ref{eq:formula}), i.e., 
\begin{equation}
\text{Tr}\left[T^{t}\left(T^{\dagger}\right)^{s},T^{m}\left(T^{\dagger}\right)^{n}\right]=\frac{{\rm i}}{2\pi}\int_{S^{1}}e^{-{\rm i}\left(t-s\right)k}\text{d}e^{-{\rm i}\left(m-n\right)k}.
\end{equation}
By straightforward calculations, we know that its left-handed side equals to $\left(m-n\right)\delta_{-t+s,m-n}$ which coincides with its right-handed side. Thus, the proof has been finished. Note that $\text{Tr}\left(XY\right)$ diverges in general case such that the cyclic property of trace does not hold and $\text{Tr}\left[X,Y\right]$ could be finite. This formula allows us to establish the BEC in a concise approach without the sophisticated \emph{K}-theory. More explicitly, if the left-handed side of Eq. (\ref{eq:formula}) depicts a topological index expressed by the edge Hamiltonian, the right-handed side corresponds to a topological invariant expressed by the bulk one.

It is worth mentioning that, to define the topological index by $H\left(\tilde{\bm{k}}\right)$, one would utilize the inverse operator of Toeplitz algebra. In Appendix \ref{sec:inverse} we show $X^{-1}\in M_{N}\left(\mathbb{C}\right)\otimes\mathbb{T}$ if $X\in M_{N}\left(\mathbb{C}\right)\otimes\mathbb{T}$ and $\varphi\left(X\right)\in M_{N}\left(\mathbb{C}\right)$ are both invertible. According to Eq. (\ref{eq:mult}), we obtain the property 
\begin{equation}
\varphi\left(X^{-1}\right)=\left[\varphi\left(X\right)\right]^{-1}.\label{eq:inverse}
\end{equation}
Remarkably, an invertible $\varphi\left(X\right)$ does not ensure $X$ invertible. Thus, the edge Hamiltonian $H\left(\tilde{\bm{k}}\right)$ could be singular despite the invertible bulk Hamiltonian $h\left(\bm{k}\right)$ which obeys the gapped condition $\det h\left(\bm{k}\right)\neq0$.

\section{BEC for odd dimensional chiral models\label{sec:chiral}}

Now let us consider the $\left(2n+1\right)$D ($n\geq0$) chiral model which belongs to class AIII in the AZ classification \citep{Chiu2016RevModPhys}. In this case, the bulk topology is characterized by a winding number with an explicit expression. After defining the edge-mode index by the edge Hamiltonian, we will prove the equivalence of these two integers without any approximation.

The bulk Hamiltonian with chiral symmetry respects a constraint 
\begin{equation}
Sh\left(\bm{k}\right)S^{-1}=-h\left(\bm{k}\right),
\end{equation}
with chiral matrix $S=I_{N/2}\oplus\left(-I_{N/2}\right)$ ($I_{m}$ denotes the identity matrix in $M_{m}\left(\mathbb{C}\right)$ and $N$ is an even number) such that it must take the off-diagonal form
\begin{equation}
h\left(\bm{k}\right)=\left(\begin{array}{cc}
0 & q\left(\bm{k}\right)\\
q^{\dagger}\left(\bm{k}\right) & 0
\end{array}\right).\label{eq:off_bulk}
\end{equation}
Due to finite hopping terms, $q\left(\boldsymbol{k}\right)$ is a smooth matrix-valued function. The gapped condition requires $\det q\left(\bm{k}\right)\neq0$ such that $q$ is a smooth mapping from $T^{2n+1}$ to the general linear group ${\rm GL}_{N/2}\left(\mathbb{C}\right)\simeq U\left(N/2\right)$ ($\simeq$ denoting the homotopy equivalence). One can define the degree of mapping as follows 
\begin{equation}
\deg=C_{n}\int_{T^{2n+1}}\text{tr}\left(q^{-1}\text{d}q\right)^{2n+1}\in\mathbb{Z},
\end{equation}
with the coefficient
\begin{equation}
C_{n}=-\frac{n!}{\left(2n+1\right)!}\left(\frac{{\rm i}}{2\pi}\right)^{n+1}.
\end{equation}
Here $\text{d}$ is the exterior differentiation and the power of $q^{-1}\text{d}q$ refers to the power of exterior product. The degree of mapping is also called as the winding number.

Like the bulk Hamiltonian, the edge Hamiltonian also takes the off-diagonal form 
\begin{equation}
H\left(\tilde{\bm{k}}\right)=\left(\begin{array}{cc}
 & Q\left(\tilde{\bm{k}}\right)\\
Q^{\dagger}\left(\tilde{\bm{k}}\right)
\end{array}\right),\label{eq:off_edge}
\end{equation}
with homomorphism $\varphi\left(Q\left(\tilde{\bm{k}}\right)\right)=q\left(\bm{k}\right)$. Despite the invertible bulk Hamiltonian, the edge Hamiltonian $H\left(\tilde{\bm{k}}\right)$ could be singular and thus have zero modes that are nontrivial bound states $\psi$ satisfying $H\psi=0$. These zero modes would exhibit a topological property reflected by an edge-mode index. This index has distinct definitions in one and higher dimensions, and hence we discuss them separately and prove the BEC respectively.

\subsection{1D case}

Due to the chirality, the zero-mode space is decomposed to a negative-chirality subspace and a positive-chirality one, i.e., $\ker H=\ker Q\oplus\ker Q^{\dagger}$ ($\ker$ denotes the kernel). And due to invertible $\varphi\left(Q\right)=q$, $Q$ is a Fredholm operator with finite dimensional $\ker Q$ and $\ker Q^{\dagger}$\,\footnote{Atkinson's theorem tells that $Q$ is a Fredholm operator if and only if $\exists P$ making $PQ-I$ and $QP-I$ compact. This can be realized by constructing a $P$ satisfying $\varphi\left(P\right)=q^{-1}$ such that $\varphi\left(PQ-I\right)=\varphi\left(QP-I\right)=0$. And Appendix \ref{sec:inverse} reveals that $\varphi\left(K\right)=0$ implies that $K$ is compact.}. Therefore, the following Fredholm index is well defined
\begin{equation}
\text{ind}=\dim\left(\ker Q\right)-\dim\left(\ker Q^{\dagger}\right)\in\mathbb{Z}
\end{equation}
and it serves as the edge-mode index for 1D chiral models. In a physical sense, its explicit expression is given by a couple of partition functions in zero temperature\,\citep{Zhou2018JPA} 
\begin{equation}
\text{ind}=\text{Tr}\lim_{\beta\rightarrow+\infty}\left(e^{-\beta Q^{\dagger}Q}-e^{-\beta QQ^{\dagger}}\right).\label{eq:ind_1D}
\end{equation}
The strict deduction resorts to the fact that Fredholm operator $Q$ restricted to $\left(\ker Q\right)^{\perp}$ ($\perp$ the orthogonal complement) has a lower bound such that $e^{-\beta Q^{\dagger}Q}\rightarrow0$, and the same for $Q^{\dagger}$.

Now we prove the BEC that says the bulk winding number equaling to the Fredholm index, i.e., $\deg=\text{ind}$. This result is known as the Toeplitz index theorem\,\citep{Bottcher2005}. Firstly, we recast the edge-mode index Eq. (\ref{eq:ind_1D}) as a trace of commutator 
\begin{equation}
\text{ind}=\text{Tr}\lim_{\beta\rightarrow+\infty}\left[\sum_{m=1}^{\infty}\frac{\left(-\beta\right)^{m}}{m!}\left(Q^{\dagger}Q\right)^{m-1}Q^{\dagger},Q\right].
\end{equation}
According to the formula Eq. (\ref{eq:formula}) and properties Eqs. (\ref{eq:linear}--\ref{eq:conju}), it reduces to 
\begin{eqnarray}
\text{ind} & = & \frac{{\rm i}}{2\pi}\lim_{\beta\rightarrow+\infty}\int_{S^{1}}\text{tr}\sum_{m=1}^{\infty}\frac{\left(-\beta\right)^{m}}{m!}\left(q^{\dagger}q\right)^{m-1}q^{\dagger}\text{d}q\\
 & = & \frac{{\rm i}}{2\pi}\lim_{\beta\rightarrow+\infty}\int_{S^{1}}\text{tr}\frac{e^{-\beta q^{\dagger}q}-I_{N/2}}{q^{\dagger}q}q^{\dagger}\text{d}q,
\end{eqnarray}
where $\varphi\left(Q\right)=q\left(k\right)$ has been used in the above. The gapped condition $\det q\neq0$ guarantees the positive definiteness of $q^{\dagger}q$ such that $\lim_{\beta\rightarrow+\infty}e^{-\beta q^{\dagger}q}=0$. Therefore, we arrive at 
\begin{equation}
\text{ind}=\frac{1}{2\pi{\rm i}}\int_{S^{1}}\text{tr}q^{-1}\text{d}q=\deg.
\end{equation}
Thus, the BEC for the 1D case has been quickly proved.

\subsection{$\left(2n+1\right)$D cases for $n\protect\geq1$}

The occurrence of zero modes in higher dimensions becomes a complicated issue. First of all, we exclude a trivial case that $H(\tilde{\bm{k}})$ has zero mode in the entire BZ $T^{2n}$. In this trivial case, the unique topological index is the Fredholm index 
\begin{equation}
\dim\left(\ker Q\right)-\dim\left(\ker Q^{\dagger}\right)=\frac{1}{2\pi{\rm i}}\int_{S^{1}}\text{tr}q^{-1}\frac{\partial q}{\partial k_{1}}\text{d}k_{1},
\end{equation}
and the BEC simply reduces to the 1D case. In fact, the Fredholm index preserves as $\tilde{\bm{k}}$ varies, for the right-handed side of the above equation is a homotopy invariant under the gap condition $\det q\neq0$. Hence, a nonzero Fredholm index necessarily results in a global zero mode. In the nontrivial case that we investigate,  there is no global zero mode such that the Fredholm index must be vanishing, which indicates $\dim\left(\ker Q\right)=\dim\left(\ker Q^{\dagger}\right)$ for all $\tilde{\bm{k}}$.

\begin{figure}
\begin{centering}
\includegraphics[width=7.5cm]{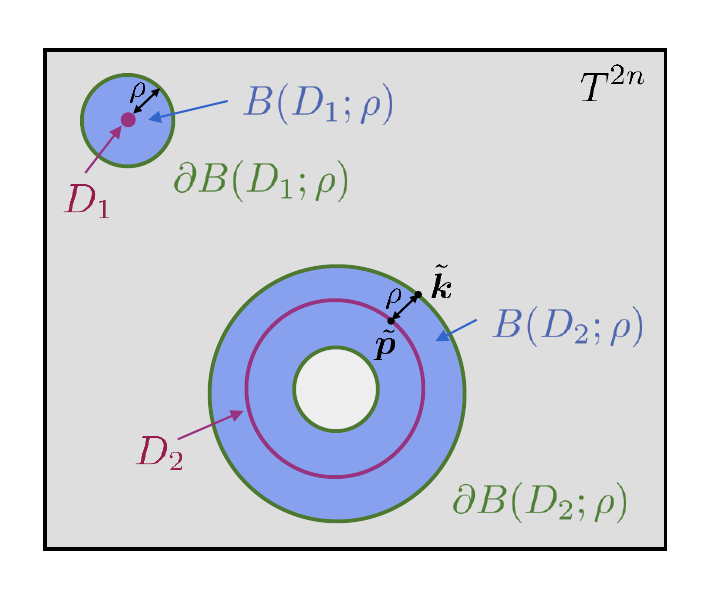} 
\par\end{centering}
\caption{\label{fig:1}Zero-mode domain of the edge Hamiltonian for $\left(2n+1\right)$D chiral models ($n\protect\geq1$). The domain of zero modes (red) in BZ $T^{2n}$ consists of smooth closed submanifolds (or discrete points) $D_{i}$ contained in a neighborhood $B\left(D_{i};\rho\right)$ (blue), with $\rho$ the distance between the boundary of neighborhood $\partial B\left(D_{i};\rho\right)$ (green) and $D_{i}$. Here $\partial B\left(D_{i};\rho\right)$ is a $\left(2n-1\right)$D closed manifold.}
\end{figure}

Since $H(\tilde{\bm{k}})$ is smooth regarding with $\tilde{\bm{k}}$, each of its zero mode forms a smooth vector field $\psi(\tilde{\bm{k}})$ whose domain is a smooth submanifold or a single point in BZ $T^{2n}$. The smooth submanifold should not have an edge and is hence closed. Moreover, the dimensions of these closed submanifolds must be lower than $2n$, for a $\left(2n\right)$D submanifold in $T^{2n}$ (but not equaling $T^{2n}$) necessarily has a boundary and is not closed. Due to probable essential degeneracy, some zero modes may share the same domain $D_{i}$. Different zero-mode domains may overlap with each other and cause accidental degeneracy, nevertheless, we can treat them separately. Eventually, the entire domain of zero modes $D=\left\{ \tilde{\bm{p}}|\ker H\left(\tilde{\bm{p}}\right)\neq0\right\} =\bigcup_{i}D_{i}$ consists of several lower-dimensional closed submanifolds (or discrete points) $D_{i}$ (see Fig. \ref{fig:1}), with each $D_{i}$ corresponding to a fixed degree of degeneracy.

Now we adopt a \emph{non-singular} hypothesis: the derivative of zero-mode spectra along the normal directions of $D_{i}$ is always nonzero. This implies the edge-mode spectra \emph{crossing }zero energy directly along normal directions such that the zero mode is substantial. In physical language, the zero mode has a linear dispersion with  finite group velocity. Physical systems (i.e., topological insulators) usually meet with this condition. Evidently, this hypothesis automatically ensures the absence of global zero mode and hence a vanishing Fredholm index. 

Then, we are able to define the edge-mode index for $\left(2n+1\right)$D chiral models ($n\geq1$). To be specific, we construct a $\left(2n\right)$D neighborhood $B\left(D_{i};\rho\right)=\bigcup_{\tilde{\bm{p}}\in D_{i}}B\left(\tilde{\bm{p}};\rho\right)$ for each $D_{i}$, in which $B\left(\tilde{\bm{p}};\rho\right)$ denotes the ball with center $\tilde{\bm{p}}$ and radius $\rho\rightarrow0$ (see Fig. \ref{fig:1}). On the boundary $\partial B\left(D_{i};\rho\right)$, we define a \emph{finite-size} matrix 
\begin{equation}
g_{i}\left(\tilde{\bm{k}}\right)=\frac{1}{\rho}\Psi_{+}^{\dagger}Q\left(\tilde{\bm{k}}\right)\Psi_{-},\label{eq:g}
\end{equation}
in which $\Psi_{+}=\left(\psi_{1},\ldots,\psi_{s}\right)$ and $\Psi_{-}=\left(\psi_{s+1},\ldots,\psi_{2s}\right)$ is the orthonormal basis of $\ker Q^{\dagger}\left(\tilde{\bm{p}}\right)$ and $\ker Q\left(\tilde{\bm{p}}\right)$ (their identical dimensions stemmed from the vanishing Fredholm index), respectively. Here $\tilde{\bm{p}}\in D_{i}$ is the nearest zero-mode point to $\tilde{\bm{k}}$. According to the perturbation theory, the derivative of zero-mode spectra along the normal direction $\tilde{\bm{k}}-\tilde{\bm{p}}$ is given by the eigenvalues of following matrix
\begin{equation}
\lim_{\rho\rightarrow0}\left(\begin{array}{c}
\Psi_{+}^{\dagger}\\
\Psi_{-}^{\dagger}
\end{array}\right)\frac{H\left(\tilde{\bm{k}}\right)-H\left(\tilde{\bm{p}}\right)}{\rho}\left(\begin{array}{cc}
\Psi_{+} & \Psi_{-}\end{array}\right)=\lim_{\rho\rightarrow0}\left(\begin{array}{cc}
 & g_{i}\left(\tilde{\bm{k}}\right)\\
g_{i}^{\dagger}\left(\tilde{\bm{k}}\right)
\end{array}\right).
\end{equation}
The non-singular hypothesis ensures $\det g_{i}\left(\tilde{\bm{k}}\right)\neq0$. Therefore, $g_{i}\left(\tilde{\bm{k}}\right)$ is a smooth mapping from the $\left(2n-1\right)$D closed manifold $\partial B\left(D_{i};\rho\right)$ to ${\rm GL}_{s}\left(\mathbb{C}\right)\simeq U\left(s\right)$, with degree of mapping 
\begin{equation}
\deg_{i}=C_{n-1}\int_{\partial B\left(D_{i};\rho\right)}\text{tr}\left(g_{i}^{-1}\text{d}g_{i}\right)^{2n-1}\in\mathbb{Z}.\label{eq:edge_wind}
\end{equation}
Note that $\deg_{i}$ is invariant under gauge transformations $\Psi_{\pm}\mapsto\Psi_{\pm}U_{\pm}$ ($U_{\pm}\in U\left(s\right)$). We call $\deg_{i}$ as the edge winding number. Finally, the overall edge-mode index is defined as a sum 
\begin{equation}
\text{ind}=\sum_{i}\deg_{i}.
\end{equation}

\begin{figure}
\begin{centering}
\includegraphics[width=7.5cm]{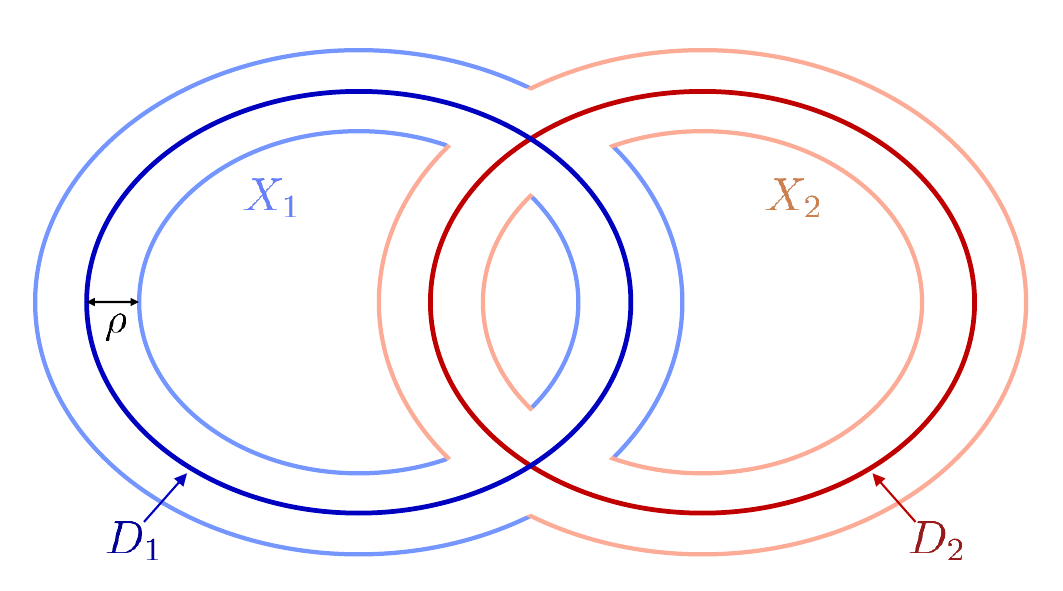}
\par\end{centering}
\caption{\label{fig:2}Treatment of Eq. (\ref{eq:ind_oddD}) for overlapping $D_{i}$. In this case, the Stokes' formula gives $C_{n-1}\text{Tr}\protect\int_{T^{2n}\backslash D}\left(Q^{-1}\text{d}Q\right)^{2n}=\sum_{i}C_{n-1}\text{Tr}\protect\int_{X_{i}}\left(Q^{-1}\text{d}Q\right)^{2n-1}$. Due to Eq. (\ref{eq:limit}), $\text{Tr}\left(Q^{-1}\text{d}Q\right)^{2n-1}=\text{tr}\left(g_{i}^{-1}\text{d}g_{i}\right)^{2n-1}$ for the most area of $X_{i}$. Besides, $X_{i}$ covers the most area of $\partial B\left(D_{i};\rho\right)$. Therefore, under the limit $\rho\rightarrow0$, $\sum_{i}C_{n-1}\text{Tr}\protect\int_{X_{i}}\left(Q^{-1}\text{d}Q\right)^{2n-1}$ could be replaced by $\sum_{i}C_{n-1}\protect\int_{\partial B\left(D_{i};\rho\right)}\text{tr}\left(g_{i}^{-1}\text{d}g_{i}\right)^{2n-1}=\sum_{i}\deg_{i}=\text{ind}$, which means that Eq. (\ref{eq:ind_oddD}) still holds.}
\end{figure}

With the edge-mode index defined, we would like to express it explicitly by the edge Hamiltonian. For simplicity, we suppose no overlap between those $D_{i}$. In Appendix \ref{sec:Q_inverse} we show 
\begin{equation}
\Psi_{-}\lim_{\rho\rightarrow0}g_{i}^{-1}\left(\tilde{\bm{k}}\right)\Psi_{+}^{\dagger}=\lim_{\rho\rightarrow0}\rho Q^{-1}\left(\tilde{\bm{k}}\right),\label{eq:limit}
\end{equation}
by which the edge-mode index can be recast as 
\begin{equation}
\text{ind}=\sum_{i}C_{n-1}\text{Tr}\lim_{\rho\rightarrow0}\int_{\partial B\left(D_{i};\rho\right)}\left(Q^{-1}\text{d}Q\right)^{2n-1}.
\end{equation}
It is worthy to emphasize that $\text{Tr}$ and $\lim$ cannot interchange, for only the limit of operator is finite-rank and hence trace-class. Then we apply Stokes' formula and achieve 
\begin{eqnarray}
\text{ind} & = & -C_{n-1}\text{Tr}\lim_{\rho\rightarrow0}\int_{T^{2n}\backslash\bigcup_{i}B\left(D_{i};\rho\right)}\text{d}\left(Q^{-1}\text{d}Q\right)^{2n-1}.\\
 & = & C_{n-1}\text{Tr}\int_{T^{2n}\backslash D}\left(Q^{-1}\text{d}Q\right)^{2n},\label{eq:ind_oddD}
\end{eqnarray}
Here $Q^{-1}(\tilde{\bm{k}})$ exists for $\tilde{\bm{k}}\in T^{2n}\backslash D$, where $H(\tilde{\bm{k}})$ has no zero mode, for $\ker Q=0$ means $Q$ injective and $\ker Q^{\dagger}=0$ means $Q$ surjective. Actually, more careful analysis shows that Eq. (\ref{eq:ind_oddD}) holds even if those $D_{i}$ overlap with each other (see Fig. \ref{fig:2}). Remarkably, the integrand belongs to $M_{N}\left(\mathbb{C}\right)\otimes\mathbb{T}$.

Now we are ready to prove the BEC that states $\deg=\text{ind}$. We still recast the index Eq. (\ref{eq:ind_oddD}) as a trace of commutator (commutator of Toeplitz algebra is trace-class) and apply the formula Eq. (\ref{eq:formula}), yielding 
\begin{eqnarray}
\text{ind} & = & C_{n-1}\text{Tr}\int_{T^{2n}\backslash D}\sum_{i=2}^{2n+1}\frac{1}{2}\left[\left(Q^{-1}\text{d}Q\right)^{2n-1},Q^{-1}\partial_{i}Q\right]\land\text{d}k_{i}\\
 & = & \frac{{\rm i}}{4\pi}C_{n-1}\int_{S^{1}}\text{d}k_{1}\int_{T^{2n}\backslash D}\text{tr}\varphi\left[\left(Q^{-1}\text{d}Q\right)^{2n-1}\right]\land\partial_{1}\varphi\left(Q^{-1}\text{d}Q\right).
\end{eqnarray}
Then we apply properties Eqs. (\ref{eq:linear}--\ref{eq:conju}) and Eq. (\ref{eq:inverse}), reducing the above equation to 
\begin{equation}
\text{ind}=\frac{{\rm i}}{4\pi}C_{n-1}\int_{T^{2n+1}}\text{d}k_{1}\land\text{tr}\left(q^{-1}\text{d}q\right)^{2n-1}\land\partial_{1}\left(q^{-1}\text{d}q\right),
\end{equation}
where $\varphi\left(Q\right)=q$ has been substituted into the equation. Here the integration area $S^{1}\times\left(T^{2n}\backslash D\right)$ has been replaced by $S^{1}\times T^{2n}=T^{2n+1}$ because $D$ has a vanishing $\left(2n\right)$D volume and $q^{-1}\left(\bm{k}\right)$ no longer has singularity in the whole BZ $T^{2n+1}$ (due to gapped condition $\det q\left(\boldsymbol{k}\right)\neq0$).

To see the final result, we continue the calculation and obtain 
\begin{eqnarray}
\text{ind} & = & \frac{{\rm i}}{4\pi}C_{n-1}\int_{T^{2n+1}}\text{d}k_{1}\land\text{tr}\left(q^{-1}\text{d}q\right)^{2n-1}\land\left(-q^{-1}\partial_{1}q\right)\cdot q^{-1}\text{d}q\nonumber \\
 &  & +\text{d}\left[\text{d}k_{1}\land\text{tr}\left(q^{-1}\text{d}q\right)^{2n-1}q^{-1}\cdot\partial_{1}q\right].
\end{eqnarray}
Here identity $\text{d}\left[\left(q^{-1}\text{d}q\right)^{2n-1}q^{-1}\right]=0$ has been used. The second term is vanishing since it integrates an exact form over a closed manifold. Finally, the edge-mode index reduces to 
\begin{eqnarray}
\text{ind} & = & \frac{{\rm i}}{4\pi}C_{n-1}\int_{T^{2n+1}}\text{tr}q^{-1}\text{d}k_{1}\partial_{1}q\land\left(q^{-1}\text{d}q\right)^{2n}\\
 & = & \frac{{\rm i}}{4\pi}C_{n-1}\frac{1}{2n+1}\int_{T^{2n+1}}\text{tr}\left(q^{-1}\text{d}q\right)^{2n+1}\\
 & = & C_{n}\int_{T^{2n+1}}\text{tr}\left(q^{-1}\text{d}q\right)^{2n+1}=\deg.
\end{eqnarray}
It is seen that the BEC for $\left(2n+1\right)$D cases has been proved. It is worthy to mention that the conclusion still holds if parameter $\tilde{\bm{k}}$ belongs to other closed manifold.

At last, we completely state our conclusions as follows.\\
\textbf{Proposition 1}: Suppose a $\left(2n+1\right)$D chiral model $h\left(\boldsymbol{k}\right)=\left(\begin{array}{cc}
 & q\\
q^{\dagger}
\end{array}\right)$ with a smooth $q\left(\boldsymbol{k}\right)$ satisfying the gapped condition $\det q\left(\boldsymbol{k}\right)\neq0$. And its corresponding edge Hamiltonian $H(\tilde{\boldsymbol{k}})=\left(\begin{array}{cc}
 & Q\\
Q^{\dagger}
\end{array}\right)$ satisfies the non-singular hypothesis (the normal derivative of zero-mode spectra at the zero-mode domain is always nonzero) if $n\geq1$. Then
\begin{equation}
\frac{1}{2\pi{\rm i}}\int_{S^{1}}\text{tr}q^{-1}\text{d}q=\dim\left(\ker Q\right)-\dim\left(\ker Q^{\dagger}\right),\,n=0,
\end{equation}
\begin{equation}
-\frac{n!}{\left(2n+1\right)!}\left(\frac{{\rm i}}{2\pi}\right)^{n+1}\int_{T^{2n+1}}\text{tr}\left(q^{-1}\text{d}q\right)^{2n+1}=\sum_{i}\deg_{i},\,n\geq1,
\end{equation}
in which $\deg_{i}$ is defined in Eq. (\ref{eq:edge_wind}).

\section{BEC for even dimensional Chern insulators\label{sec:Chern insulator}}

In this section we consider the $\left(2n\right)$D ($n\geq1$) Chern insulators without any internal symmetry, thus belonging to class A in the AZ classification. In this case, the bulk topology of Hamiltonians is characterized by the Chern number of the valence (negative-energy) bands which is formulated by Berry curvature (see Appendix \ref{sec:Chern}). To apply our formula, we need to explicitly write the Chern number by bulk Hamiltonian. This could be done by a simple construction.

Actually, every $\left(2n\right)$D Chern insulator Hamiltonian $h\left(\bm{k}\right)$ can be mapped to a $\left(2n+1\right)$D chiral model in form of Eq. (\ref{eq:off_bulk}) via construction 
\begin{equation}
q\left(\bm{k},\omega\right)=h\left(\bm{k}\right)-{\rm i}\omega I_{N}
\end{equation}
with $\omega\in\mathbb{R}\cup\left\{ \infty\right\} \simeq S^{1}$. The positive definiteness of $q^{\dagger}q$ guarantees the gapped condition $\det q\left(\bm{k},\omega\right)\neq0$ such that the degree of mapping $q$ is well defined, i.e., 
\begin{equation}
\deg=C_{n}\int_{T^{2n}\times S^{1}}\text{tr}\left(q^{-1}\text{d}q\right)^{2n+1}.\label{eq:const}
\end{equation}
In Appendix \ref{sec:Chern}, we show that the Chern number ${\rm Ch}$ of the original Hamiltonian is given by 
\begin{equation}
{\rm Ch}=\left(-1\right)^{n}\deg.\label{eq:deg=Ch}
\end{equation}
In this way, the Chern number has an explicit expression regarding with the bulk Hamiltonian $h\left(\bm{k}\right)$.

Similarly, the edge Hamiltonian $H\left(\tilde{\bm{k}}\right)$ can also be mapped to the chiral model Eq. (\ref{eq:off_edge}) with 
\begin{equation}
Q\left(\tilde{\bm{k}},\omega\right)=H\left(\tilde{\bm{k}}\right)-{\rm i}\omega I,
\end{equation}
and the definition of the edge-mode index for the Chern insulator can be induced by that of the chiral model. Like the previous treatment, we still adopt the non-singular hypothesis that the spectra crosses zero energy substantially, which automatically ensures the absence of global zero mode. Then the zero-mode domain $D=\left\{ \tilde{\bm{p}}|\ker H\left(\tilde{\bm{p}}\right)\neq0\right\} \subset T^{2n-1}$ consists of several lower-dimensional closed manifolds (or discrete points) $D_{i}$, with each $D_{i}$ corresponding to a fixed degree of degeneracy $\dim\left(\ker H\right)$. 

\begin{figure}
\begin{centering}
\includegraphics[width=7.5cm]{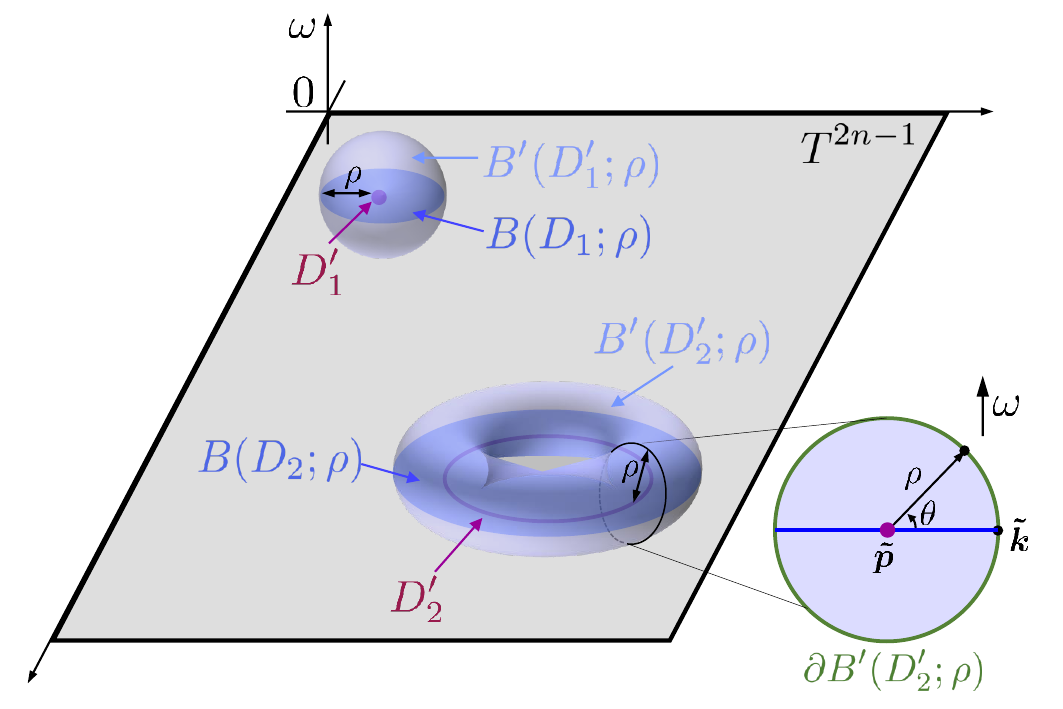} 
\end{centering}
\caption{\label{fig:3}Zero-mode domains of the edge Hamiltonian for $\left(2n\right)$D Chern insulators and the constructed $\left(2n+1\right)$D chiral models. The zero-mode domain of $H(\tilde{\bm{k}})$ in BZ $T^{2n-1}$ consists of smooth closed manifolds $D_{i}$ (red) contained in a $\left(2n-1\right)$D neighborhood $B\left(D_{i};\rho\right)$ (dark blue). Correspondingly the zero-mode domain of $Q(\tilde{\bm{k}},\omega)=H(\tilde{\bm{k}})-{\rm i}\omega I$ in $T^{2n-1}\times S^{1}$ consists of $D_{i}^{\prime}=\left(D_{i},0\right)$ contained in a $\left(2n\right)$D neighborhood $B^{\prime}\left(D_{i}^{\prime};\rho\right)$ (light blue). Here $\partial B\left(D_{i};\rho\right)$ and $\partial B^{\prime}\left(D_{i};\rho\right)$ are a $\left(2n-2\right)$D and a $\left(2n-1\right)$D closed manifold, respectively.}
\end{figure}

Correspondingly, the zero-mode domain of $Q(\tilde{\bm{k}},\omega)$ consists of $D_{i}^{\prime}=\left(D_{i},0\right)$ (abbreviation of $\left\{ \left(\tilde{\bm{p}},0\right)|\tilde{\bm{p}}\in D_{i}\right\} $). Each $D_{i}^{\prime}$ has a $\left(2n\right)$D neighborhood $B^{\prime}\left(D_{i}^{\prime};\rho\right)=\left\{ \left(B\left(D_{i};\rho\cos\theta\right),\rho\sin\theta\right)|\theta\in\left[-\frac{\pi}{2},\frac{\pi}{2}\right]\right\} $ (see Fig. \ref{fig:3}). Each point on $\partial B^{\prime}\left(D_{i}^{\prime};\rho\right)$ can be expressed as $(\tilde{\bm{p}}+(\tilde{\bm{k}}-\tilde{\bm{p}})\cos\theta,\rho\sin\theta)$ with a certain $\tilde{\bm{k}}\in\partial B\left(D_{i};\rho\right)$ and the nearest zero-mode point $\tilde{\bm{p}}\in D_{i}$. Then the finite-size matrix $g_{i}$ defined by Eq. (\ref{eq:g}) reads (up to the main order of $\rho$)

\begin{eqnarray}
g_{i} & = & \frac{1}{\rho}\Psi^{\dagger}\left[\left(\tilde{\bm{k}}-\tilde{\bm{p}}\right)\cos\theta\cdot\nabla H\left(\tilde{\bm{p}}\right)\right]\Psi-{\rm i}I_{s}\sin\theta\\
 & = & \frac{1}{\rho}\Psi^{\dagger}H\left(\tilde{\bm{k}}\right)\Psi\cos\theta-{\rm i}I_{s}\sin\theta,
\end{eqnarray}
in which $\Psi=\left(\psi_{1},\cdots,\psi_{s}\right)$ is the orthonormal basis of $\ker H\left(\tilde{\bm{p}}\right)$. Therefore, the degree of mapping $g:\partial B^{\prime}\left(D_{i}^{\prime};\rho\right)\rightarrow{\rm GL}_{s}\left(\mathbb{C}\right)\simeq U\left(s\right)$ is given by 
\begin{eqnarray}
\deg_{i} & = & C_{n-1}\int_{\partial B^{\prime}\left(D_{i}^{\prime};\rho\right)}\text{tr}\left(g_{i}^{-1}\text{d}g_{i}\right)^{2n-1}\\
 & = & C_{n-1}\int_{\partial B\left(D_{i};\rho\right)\times\mathbb{R}}\text{tr}\left[\frac{1}{G_{i}-{\rm i}\nu I_{s}}\text{d}\left(G_{i}-{\rm i}\nu I_{s}\right)\right]^{2n-1},\label{eq:deg_i}
\end{eqnarray}
with $\nu=\tan\theta\in\mathbb{R}$, and the reduced Hamiltonian on $\partial B\left(D_{i};\rho\right)$ is defined by
\begin{equation}
G_{i}\left(\tilde{\bm{k}}\right)=\frac{1}{\rho}\Psi^{\dagger}H\left(\tilde{\bm{k}}\right)\Psi.\label{eq:red_Hamil}
\end{equation}
The non-singular hypothesis implies $\det G_{i}(\tilde{\bm{k}})\neq0$, i.e., the reduced Hamiltonian is still gapped.

Now the edge-mode index of the Chern insulator can be induced. For $n\geq2$, we compare Eq. (\ref{eq:deg_i}) with Eqs. (\ref{eq:const}--\ref{eq:deg=Ch}) and conclude that 
\[
\deg_{i}=\left(-1\right)^{n-1}{\rm Ch}_{i},\,n\geq2,
\]
in which ${\rm Ch}_{i}$ is the Chern number for the ``valence bands'' of reduced Hamiltonian $G_{i}(\tilde{\bm{k}})$ over $\left(2n-2\right)$D closed manifold $\partial B\left(D_{i};\rho\right)$, dubbed as edge Chern number. For $n=1$, the parameter $\tilde{k}$ is 1D and the zero-mode domain $D_{i}$ must be a single point $\tilde{p}$. Then $\partial B\left(D_{i};\rho\right)=\left\{ \tilde{p}+\rho\right\} -\left\{ \tilde{p}-\rho\right\} \simeq S^{0}$ ($\left\{ \tilde{p}+\rho\right\} $ and $-\left\{ \tilde{p}-\rho\right\} $ refer to 0-chain whose coefficients reflect the orientations) such that $\deg_{i}$ reduces to 
\begin{eqnarray}
\deg_{i} & = & \frac{1}{2\pi{\rm i}}\int_{-\infty}^{+\infty}\text{tr}\left[\frac{-{\rm i}\text{d}\nu}{G_{i}\left(\tilde{k}\right)-{\rm i}\nu I}\right]_{\tilde{k}=\tilde{p}-\rho}^{\tilde{p}+\rho}\\
 & = & -\text{tr}\,\text{sgn}\left[G_{i}\left(\tilde{p}+\rho\right)-G_{i}\left(\tilde{p}-\rho\right)\right]\\
 & = & -\text{tr}\,\text{sgn}G_{i}^{\prime}\left(\tilde{p}\right).
\end{eqnarray}
Here $\text{tr}\,\text{sgn}G_{i}^{\prime}\left(\tilde{p}\right)$ (the prime denotes the derivative) is the spectral flow of $H\left(k_{2}\right)$ at $k_{2}=\tilde{p}$, which counts the sign flipping of edge-mode spectra when $k_{2}$ passes through $\tilde{p}$. The negative sum of edge Chern number $-\sum_{i}{\rm Ch}_{i}$ or the total spectral flow $\sum_{i}\text{tr}\,\text{sgn}G_{i}^{\prime}$ serves as the edge-mode index.

Finally, we recall the BEC for the chiral models and achieve 
\begin{equation}
\left(-1\right)^{n}{\rm Ch}=\sum_{i}\deg_{i},
\end{equation}
which can be recast as 
\begin{eqnarray}
{\rm Ch} & = & -\sum_{i}{\rm Ch}_{i},\,n\geq2,\\
{\rm Ch} & = & \sum_{i}\text{tr}\,\text{sgn}G_{i}^{\prime},\,n=1.
\end{eqnarray}
These two equations reflect the BEC of Chern insulator. Remarkably, the last equation reproduces the well known conclusion of the 2D Chern insulator, that its Chern number counts the overall sign flipping of the edge-mode energy.

The complete conclusions are stated as follows.\\
\textbf{Proposition 2}: Suppose a $\left(2n\right)$D Chern insulator described by a smooth Hamiltonian $h\left(\boldsymbol{k}\right)$ that satisfies gapped condition $\det h\left(\boldsymbol{k}\right)\neq0$. And its corresponding edge Hamiltonian $H(\tilde{\boldsymbol{k}})$ satisfies the non-singular hypothesis. Then one can define several $\left(2n-2\right)$D gapped Hamiltonians $G_{i}(\tilde{\bm{k}})$ near the zero-mode domains according to Eq. (\ref{eq:red_Hamil}) such that
\begin{eqnarray}
{\rm Ch} & = & \sum_{i}\text{tr}\,\text{sgn}G_{i}^{\prime},\,n=1,\\
{\rm Ch} & = & -\sum_{i}{\rm Ch}_{i},\,n\geq2,
\end{eqnarray}
in which ${\rm Ch}$ and ${\rm Ch}_{i}$ is the Chern number of the valence bands for $h\left(\boldsymbol{k}\right)$ and $G_{i}(\tilde{\bm{k}})$, respectively.

\section{Conclusions\label{sec:conclusions}}

In conclusion, the BEC for the complex AZ classes in any dimension has been completely proved. The key step of the proof is the formula of Toeplitz algebra that links the quantity expressed by edge Hamiltonian to that by bulk one. It enables us to rigorously prove the BEC by calculus computations after the bulk and edge topological quantities are explicitly expressed by Hamiltonians.

For odd-dimensional chiral models, the bulk winding number already has an explicit expression. In the 1D case, the edge-mode index is the Fredholm index of the edge Hamiltonian. With the help of our formula, the BEC, that the winding number equals to the Fredhom index, has been quickly proved. In higher dimensional cases, the edge-mode index has been defined as the sum of edge winding numbers, and it has also been proved to be identical to the bulk winding number.

For even-dimensional Chern insulators, after they are mapped to the odd-dimensional chiral models, the bulk Chern number and edge-mode index are both induced by the topological quantities of the constructed chiral model. Hence, the BEC of the Chern insulators are inherited from that of the chiral models. It turns out that the bulk Chern number equals to the spectral flow (for 2D case) or the negative sum of edge Chern numbers (for higher-even dimensions).

Our work provides a clear understanding to the BEC without the sophisticated \emph{K}-theory and can probably be extended to other physical scenarios like the BEC for $\mathbb{Z}_{2}$ topological insulators in real AZ classes or the BEC for topological Bogoliubov excitation of bosons.

\section*{Acknowledgements}
We thank Prof. Zhi-Fang Xu for his helpful discussions. This work is supported by the Natural Science Foundation of Top Talent of SZTU (GDRC202202, GDRC202312), Guangdong Provincial Quantum Science Strategic Initiative (No. GDZX2305006), and Characteristic Innovation Project of Guangdong Provincial Universities (No. 2024KTSCX053).

\appendix

\section{Inverse Operator of Toeplitz algebra \label{sec:inverse}}

Here we prove $X^{-1}\in M_{N}\left(\mathbb{C}\right)\otimes\mathbb{T}$ if $X\in M_{N}\left(\mathbb{C}\right)\otimes\mathbb{T}$ and $\varphi\left(X\right)\in M_{N}\left(\mathbb{C}\right)$ are both invertible. (Actually, an invertible $X$ necessarily leads to an invertible $\varphi\left(X\right)$, however, it costs additional proof and we do not need such strong conclusion here.)

Firstly, any element in $M_{N}\left(\mathbb{C}\right)\otimes\mathbb{T}$, expressed by $X=\sum_{m,n\in\mathbb{N}}x_{mn}\otimes T^{m}\left(T^{\dagger}\right)^{n}$ with $x_{mn}\in M_{N}\left(\mathbb{C}\right)$, can be decomposed as $X=T_{\varphi\left(X\right)}+K$. Here the Toeplitz operator $T_{f}$ with continuous symbol $f\left(k\right)=\sum_{l\in\mathbb{Z}}f_{l}e^{-{\rm i}lk}$ ($f_{l}\in M_{N}\left(\mathbb{C}\right)$) is defined as  
\begin{equation}
T_{f}=\sum_{l\geq0}f_{l}T^{l}+\sum_{l>0}f_{-l}\left(T^{\dagger}\right)^{l}
\end{equation}
with property $\varphi\left(T_{f}\right)=f$. And the operator 
\begin{equation}
K=\sum_{l=0}^{\infty}\sum_{n}x_{n+l,n}\otimes T^{l}\left[T^{n}\left(T^{\dagger}\right)^{n}-I\right]+\sum_{l=1}^{\infty}\sum_{m}x_{m,m+l}\otimes\left[T^{m}\left(T^{\dagger}\right)^{m}-1\right]\left(T^{\dagger}\right)^{l}
\end{equation}
is a limit of finite-rank matrix and hence compact, for $T^{l}\left[T^{n}\left(T^{\dagger}\right)^{n}-I\right]$ and $\left[T^{m}\left(T^{\dagger}\right)^{m}-1\right]\left(T^{\dagger}\right)^{l}$ are both finite-rank matrices. Evidently, $\varphi\left(K\right)=0$. This decomposition leads to a corollary: $X\in M_{N}\left(\mathbb{C}\right)\otimes\mathbb{T}$ is compact if and only if $\varphi\left(X\right)=0$.

Next, the invertible $X$ and $\varphi\left(X\right)$ allow us to construct operator 
\begin{equation}
O=X^{-1}-T_{\left[\varphi\left(X\right)\right]^{-1}}.
\end{equation}
Since $XO=I-XT_{\left[\varphi\left(X\right)\right]^{-1}}\in M_{N}\left(\mathbb{C}\right)\otimes\mathbb{T}$,
we are able to compute 
\begin{equation}
\varphi\left(XO\right)=\varphi\left(I-XT_{\left[\varphi\left(X\right)\right]^{-1}}\right)=0,
\end{equation}
which implies $XO$ being compact. On the other hand, since $X$ is a bounded invertible operator, its inverse $X^{-1}$ is also bounded. It turns out that $O$ is compact due to the fact that a bounded operator multiplying a compact operator produces a compact one, i.e., $X^{-1}\cdot XO=O$.

In general, the compact operator $O$ can be expanded by finite-rank matrices, i.e., 
\begin{equation}
O=\sum_{m,n\in\mathbb{N}^{+}}o_{mn}\otimes J_{mn},\,o_{mn}\in M_{N}\left(\mathbb{C}\right),
\end{equation}
where $J_{mn}\in M_{\infty}\left(\mathbb{C}\right)$ is the matrix whose entry at row $m$ and column $n$ is $1$, with all the other entries vanishing. Each $J_{mn}$ can be expressed by Toeplitz algebra, i.e., 
\begin{eqnarray}
J_{mm} & = & T^{m-1}\left(T^{\dagger}\right)^{m-1}-T^{m}\left(T^{\dagger}\right)^{m}\in\mathbb{T},\,m\in\mathbb{N}^{+},\\
J_{m,m+l} & = & J_{mm}\left(T^{\dagger}\right)^{l}\in\mathbb{T},\,l\in\mathbb{N}^{+},\\
J_{m+l,m} & = & T^{l}J_{mm}\in\mathbb{T},\,l\in\mathbb{N}^{+}.
\end{eqnarray}
Therefore, $O\in M_{N}\left(\mathbb{C}\right)\otimes\mathbb{T}$.

Eventually, the bounded operator $X^{-1}=T_{\left[\varphi\left(X\right)\right]^{-1}}+O$ has been expressed by $M_{N}\left(\mathbb{C}\right)\otimes\mathbb{T}$, i.e., $X^{-1}\in M_{N}\left(\mathbb{C}\right)\otimes\mathbb{T}$.

\section{$Q^{-1}$ near zero-mode point \label{sec:Q_inverse}}

We compute $Q^{-1}\left(\tilde{\bm{k}}\right)$ in the limit $\rho=\left\Vert \tilde{\bm{k}}-\tilde{\bm{p}}\right\Vert \rightarrow0$ in which $\tilde{\bm{p}}$ is a zero-mode point. Our main idea is to solve linear equation $Q\left(\tilde{\boldsymbol{p}}\right)\psi+\left[Q\left(\tilde{\bm{k}}\right)-Q\left(\tilde{\bm{p}}\right)\right]\psi=\phi$ by perturbation theory where $\left[Q\left(\tilde{\bm{k}}\right)-Q\left(\tilde{\bm{p}}\right)\right]\propto\rho$ is infinitesimal. The solution of $\psi$ would be an infinite series whose convergence would be verified by functional analysis. The main order term of $\psi$ simply gives the limit of $Q^{-1}\left(\tilde{\bm{k}}\right)$. We denote the orthonormal basis of $\ker Q^{\dagger}\left(\tilde{\bm{p}}\right)$ and $\ker Q\left(\tilde{\bm{p}}\right)$ as $\Psi_{+}$ and $\Psi_{-}$, respectively. Note that $\ker Q^{\dagger}\left(\tilde{\bm{p}}\right)$ and $\ker Q\left(\tilde{\bm{p}}\right)$ share the same dimension $s$ due to the vanishing Fredholm index.

We firstly prove that $Q^{-1}\left(\tilde{\bm{k}}\right)$ is bounded as its domain restricted to $\text{im}Q\left(\tilde{\bm{p}}\right)$. For a vector $\phi_{0}\in\text{im}Q\left(\tilde{\bm{p}}\right)=\left[\ker Q^{\dagger}\left(\tilde{\bm{p}}\right)\right]^{\perp}$ ($\text{im}$ denoting image and $\perp$ denoting orthogonal complement), there always exists a vector $\xi_{0}$ to make $Q\left(\tilde{\bm{p}}\right)\xi_{0}=\phi_{0}$ and hence $\Psi_{+}^{\dagger}\phi_{0}=0$. The Fredholm operator $Q\left(\tilde{\bm{p}}\right)$ restricted to $\left[\ker Q\left(\tilde{\bm{p}}\right)\right]^{\perp}$ has a lower bound $L$ such that $\left\Vert Q\left(\tilde{\bm{p}}\right)\xi_{0}\right\Vert >L\left\Vert \xi_{0}\right\Vert $ or, equivalently, $\left\Vert \xi_{0}\right\Vert <L^{-1}\left\Vert \phi_{0}\right\Vert $. Based on $\xi_{0}$, we construct a new vector 
\begin{equation}
\phi_{1}=-\frac{Q\left(\tilde{\bm{k}}\right)-Q\left(\tilde{\bm{p}}\right)}{\rho}\left(\xi_{0}+\Psi_{-}v_{0}\right)\label{eq:phi1}
\end{equation}
with 
\begin{equation}
v_{0}=-g^{-1}\left(\tilde{\bm{k}}\right)\frac{1}{\rho}\Psi_{+}^{\dagger}Q\left(\tilde{\bm{k}}\right)\xi_{0}\in\mathbb{C}^{s}
\end{equation}
such that $\phi_{1}\in\left[\ker Q^{\dagger}\left(\tilde{\bm{p}}\right)\right]^{\perp}=\text{im}Q\left(\tilde{\bm{p}}\right)$,
i.e., 
\begin{equation}
\Psi_{+}^{\dagger}\phi_{1}=-\frac{1}{\rho}\Psi_{+}^{\dagger}Q\left(\tilde{\bm{k}}\right)\left(\xi_{0}+\Psi_{-}v_{0}\right)=0.
\end{equation}
Here $\frac{1}{\rho}\Psi_{+}^{\dagger}Q\left(\tilde{\bm{k}}\right)\Psi_{-}=g\left(\tilde{\bm{k}}\right)$ has been used (the subscript of $g$ has been omitted). All the operators above are bounded, therefore, there exists a bound $M_{0}$ such that $\left\Vert \Psi_{-}v_{0}\right\Vert <M_{0}\left\Vert \xi_{0}\right\Vert $ and a bound $M_{1}$ such that $\left\Vert \phi_{1}\right\Vert <M_{1}\left\Vert \xi_{0}\right\Vert $.

Now we recast Eq. (\ref{eq:phi1}) as 
\begin{equation}
Q^{-1}\left(\tilde{\bm{k}}\right)\phi_{0}=\left(\xi_{0}+\Psi_{-}v_{0}\right)+\rho Q^{-1}\left(\tilde{\bm{k}}\right)\phi_{1},
\end{equation}
and find 
\begin{eqnarray}
\left\Vert \xi_{0}+\Psi_{-}v_{0}\right\Vert  & < & L^{-1}\left(1+M_{0}\right)\left\Vert \phi_{0}\right\Vert ,\\
\left\Vert \phi_{1}\right\Vert  & < & L^{-1}M_{1}\left\Vert \phi_{0}\right\Vert .
\end{eqnarray}
Due to $\phi_{1}\in\text{im}Q\left(\tilde{\bm{p}}\right)$ again, we can continuously repeat the above treatment and obtain the expansion
\begin{equation}
Q^{-1}\left(\tilde{\bm{k}}\right)\phi_{0}=\left(\xi_{0}+\Psi_{-}v_{0}\right)+\rho\left(\psi_{1}+\Psi_{-}v_{1}\right)+\cdots,
\end{equation}
with 
\begin{eqnarray}
\left\Vert Q^{-1}\left(\tilde{\bm{k}}\right)\phi_{0}\right\Vert  & < & L^{-1}\left(1+M_{0}\right)\left\Vert \phi_{0}\right\Vert +L^{-1}\left(1+M_{0}\right)\cdot\rho L^{-1}M_{1}\left\Vert \phi_{0}\right\Vert +\cdots\\
 & = & \frac{L^{-1}\left(1+M_{0}\right)}{1-\rho L^{-1}M_{1}}\left\Vert \phi_{0}\right\Vert .
\end{eqnarray}
Thus $Q^{-1}\left(\tilde{\bm{k}}\right)$ restricted to $\text{im}Q\left(\tilde{\bm{p}}\right)$ is bounded, which indicates 
\begin{equation}
\lim_{\rho\rightarrow0}\rho Q^{-1}\left(\tilde{\bm{k}}\right)\phi_{0}=0,\,\forall\phi_{0}\in\text{im}Q\left(\tilde{\bm{p}}\right).\label{eq:imQ}
\end{equation}

We next calculate $Q^{-1}\left(\tilde{\bm{k}}\right)$ restricted to $\left[\text{im}Q\left(\tilde{\bm{p}}\right)\right]^{\perp}$. A vector $\phi\in\left[\text{im}Q\left(\tilde{\bm{p}}\right)\right]^{\perp}=\ker Q^{\dagger}\left(\tilde{\bm{p}}\right)$ can always be expanded as $\phi=\Psi_{+}v$ with $v\in\mathbb{C}^{s}$. And we construct 
\begin{equation}
\phi_{0}=\phi-\frac{1}{\rho}Q\left(\tilde{\bm{k}}\right)\Psi_{-}g^{-1}\left(\tilde{\bm{k}}\right)v
\end{equation}
which satisfies $\Psi_{+}^{\dagger}\phi_{0}=0$, i.e., $\phi_{0}\in\text{im}Q\left(\tilde{\bm{p}}\right)$. From the previous result, we know $\lim_{\rho\rightarrow0}\rho Q^{-1}\left(\tilde{\bm{k}}\right)\phi_{0}=0$. Therefore, we obtain 
\begin{eqnarray}
\lim_{\rho\rightarrow0}\rho Q^{-1}\left(\tilde{\bm{k}}\right)\phi & = & \lim_{\rho\rightarrow0}\rho Q^{-1}\left(\tilde{\bm{k}}\right)\phi_{0}+\lim_{\rho\rightarrow0}\Psi_{-}g^{-1}\left(\tilde{\bm{k}}\right)v\\
 & = & \Psi_{-}\lim_{\rho\rightarrow0}g^{-1}\left(\tilde{\bm{k}}\right)\Psi_{+}^{\dagger}\phi.\label{eq:kerQdagger}
\end{eqnarray}
At last, we combine Eqs. (\ref{eq:imQ}) and (\ref{eq:kerQdagger}) together, and arrive at 
\begin{equation}
\lim_{\rho\rightarrow0}\rho Q^{-1}\left(\tilde{\bm{k}}\right)=\Psi_{-}\lim_{\rho\rightarrow0}g^{-1}\left(\tilde{\bm{k}}\right)\Psi_{+}^{\dagger}.
\end{equation}

\section{Expressions of Chern number \label{sec:Chern}}

The Chern number of a $\left(2n\right)$D Chern insulator is defined as follows. Firstly, we make local unitary diagonalization for the bulk Hamiltonian 
\begin{equation}
U^{\dagger}\left(\bm{k}\right)h\left(\bm{k}\right)U\left(\bm{k}\right)=E\left(\bm{k}\right),
\end{equation}
where $E\left(\bm{k}\right)$ is the diagonal matrix of eigen-energies and $U\left(\bm{k}\right)$ is unitary. Let $E=E_{+}\oplus E_{-}$ in which $E_{\pm}$ corresponds to the conductive (positive) and valence (negative) band energies, respectively. Simultaneously, the unitary matrix takes the form of $U=\left(u_{+},u_{-}\right)$ in which $u_{\pm}\left(\bm{k}\right)$ is the orthonormal basis of conductive and valence bands, respectively. Next, we consider the valence bands as a vector bundle and define its Berry connection (generally non-Abelian) as 
\begin{equation}
\mathcal{A}\left(\bm{k}\right)=u_{-}^{\dagger}\left(\bm{k}\right)\text{d}u_{-}\left(\bm{k}\right).
\end{equation}
It induces the Berry curvature 2-form 
\begin{equation}
{\cal F}=\text{d}\mathcal{A}+\mathcal{A}^{2}=-u_{-}^{\dagger}\text{d}u_{+}\land u_{+}^{\dagger}\text{d}u_{-},
\end{equation}
in which identity $\text{d}\left(U^{-1}\text{d}U\right)=-\left(U^{-1}\text{d}U\right)^{2}$ has been used. Finally, the Chern number is defined by 
\begin{eqnarray}
{\rm Ch} & = & \frac{1}{n!}\int_{T^{2n}}\text{tr}\left(\frac{{\rm i}}{2\pi}\mathcal{F}\right)^{n}\\
 & = & \frac{1}{n!}\left(\frac{{\rm i}}{2\pi}\right)^{n}\int_{T^{2n}}\text{tr}\left(-\text{d}u_{+}\cdot u_{+}^{\dagger}\land\text{d}u_{-}\cdot u_{-}^{\dagger}\right)^{n}.
\end{eqnarray}

Now we show that the Chern number is related to the winding number of the constructed chiral model given by Eq. (\ref{eq:const}). We note that the winding number is invariant under homotopy $q_{t}\left(\bm{k}\right)$ ($t\in\left[0,1\right]$) that keeps the gap condition $\det q_{t}\left(\bm{k}\right)\neq0$, for 
\begin{eqnarray}
\frac{\delta}{\delta t}\deg & = & \left(2n+1\right)C_{n}\int_{T^{2n+1}}\text{tr}\frac{\delta}{\delta t}\left(q_{t}^{-1}\text{d}q_{t}\right)\land\left(q_{t}^{-1}\text{d}q_{t}\right)^{2n}\\
 & = & \left(2n+1\right)C_{n}\int_{T^{2n+1}}\text{d}\text{tr}\left[q_{t}^{-1}\frac{\delta}{\delta t}q_{t}\cdot\left(q_{t}^{-1}\text{d}q_{t}\right)^{2n}\right]\\
 & = & 0,
\end{eqnarray}
where identity $\text{d}\left(q_{t}^{-1}\text{d}q_{t}\right)^{2n}=0$ has been used. This allows us to utilize the flattened Hamiltonian $h_{1}=U\left(\text{sgn}E\right)U^{\dagger}$ obtained from the homotopy
\begin{equation}
h_{t}\left(\bm{k}\right)=\left[h^{2}\left(\bm{k}\right)\right]^{-\frac{t}{2}}h\left(\bm{k}\right),\,t\in\left[0,1\right]
\end{equation}
to calculate Eq. (\ref{eq:const}), leading to 
\begin{equation}
\deg=\left(2n+1\right)C_{n}\int_{T^{2n+1}}\text{tr}\left(\frac{1}{\text{sgn}E-{\rm i}\omega I_{N}}\left[U^{-1}\text{d}U,\text{sgn}E\right]\right)^{2n}\land\frac{-{\rm i}\text{d}\omega}{\text{sgn}E-{\rm i}\omega I_{N}}.
\end{equation}
By substituting $U=\left(u_{+},u_{-}\right)$ into it, we reduce it to 
\begin{eqnarray}
\deg & = & -{\rm i}C_{n}\left(2n+1\right)\int\text{d}\omega\int_{T^{2n}}\left(\frac{4}{1+\omega^{2}}\right)^{n}\text{tr}\left(u_{+}^{\dagger}\text{d}u_{-}\land u_{-}^{\dagger}\text{d}u_{+}\right)^{n}\frac{1}{1-{\rm i}\omega}\nonumber \\
 &  & +\left(\frac{4}{1+\omega^{2}}\right)^{n}\text{tr}\left(u_{-}^{\dagger}\text{d}u_{+}\land u_{+}^{\dagger}\text{d}u_{-}\right)^{n}\frac{1}{-1-{\rm i}\omega}.
\end{eqnarray}
After integrating over $\omega$, we finally achieve 
\begin{eqnarray}
\deg & = & \frac{1}{n!}\left(\frac{{\rm i}}{2\pi}\right)^{n}\int_{T^{2n}}\text{tr}\left(\text{d}u_{+}\cdot u_{+}^{\dagger}\land\text{d}u_{-}\cdot u_{-}^{\dagger}\right)^{n}\\
 & = & \left(-1\right)^{n}{\rm Ch}.
\end{eqnarray}

\end{document}